\documentstyle[twoside,fleqn,espcrc2]{article}

\newcommand{\nc}{\newcommand}
\nc{\beq}{\begin{equation}} \nc{\eeq}{\end{equation}}
\nc{\beqa}{\begin{eqnarray}} \nc{\eeqa}{\end{eqnarray}}
\nc{\lsim}{\begin{array}{c}\,\sim\vspace{-21pt}\\< \end{array}}
\nc{\gsim}{\begin{array}{c}\sim\vspace{-21pt}\\> \end{array}}

\newcommand{\drawsquare}[2]{\hbox{%
\rule{#2pt}{#1pt}\hskip-#2pt
\rule{#1pt}{#2pt}\hskip-#1pt
\rule[#1pt]{#1pt}{#2pt}}\rule[#1pt]{#2pt}{#2pt}\hskip-#2pt
\rule{#2pt}{#1pt}}

\newcommand{\Yfund}{\raisebox{-.5pt}{\drawsquare{6.5}{0.4}}}

\title{Gauge-Mediated Supersymmetry Breaking within the Dynamical Messenger 
Sector\thanks{This contribution is based on an invited 
talk given by E. Poppitz at the Fifth International Conference on
Supersymmetries
in Physics, SUSY-97, Philadelphia, PA, May 27-31, 1997; Preprint EFI-97-32.}}

\author{Erich Poppitz\address{Enrico Fermi Institute, University of Chicago, 
5640 S. Ellis Ave., 
Chicago IL 60657, USA}
\thanks{Present address: 
Department of Physics, 
University of California at San Diego, 9500 Gilman Drive,  
La Jolla, CA 92093, USA.}%
        and 
        Sandip P. Trivedi\address{Fermi National Accelerator Laboratory, P.O. Box 500,
        Batavia IL 60510, USA}}
       
\begin{document}

\begin{abstract}
We consider the idea of combining the supersymmetry breaking and 
messenger sectors in models of gauge-mediated supersymmetry breaking.
We discuss the advantages and problems of such models, and 
present an existence proof by constructing an explicit example. 
We explore in some detail the low-energy supersymmetry-breaking 
dynamics  and 
the generation of supersymmetry-breaking soft masses
 of the standard model superpartners, and discuss many features
 which are likely to be shared by more realistic  models with
 dynamical messengers.
\end{abstract}

\maketitle

\section{Why dynamical messengers?}

The previous speakers in today's afternoon session discussed in detail 
the reasons for the current interest in 
gauge-mediated models of supersymmetry breaking. 
We refer the reader to the
  contributions of A. Nelson and 
C. Kolda to this volume for an exhaustive discussion and an 
extensive list of references. 

We begin by giving a definition of what we mean by
dynamical messengers. Loosely speaking, we will
call the messenger quarks and leptons dynamical
if they are  an integral part of the supersymmetry breaking sector. In a more
precise sense, we require that they carry quantum numbers under both the 
standard model gauge group, $G_{SM}$, and the gauge group responsible for
the breaking of supersymmetry, $G_{SB}$.

\subsection{The problems one wants to solve.}

\smallskip

To motivate our interest in models with dynamical messengers---in
the above well-defined sense---let us recall some features and  
problems of the original  models of minimal gauge-mediation (MGM) 
 of Dine, Nelson, and
Shirman \cite{dns}:
\begin{enumerate}
\item{
In the MGM models, the supersymmetry breaking sector
is a ``black box"---its only relevant feature  is that it gives rise
to both a supersymmetry preserving and a 
supersymmetry breaking vacuum expectation value of 
a standard model singlet chiral superfield---the so-called  
``messenger singlet". To realize this 
feature dynamically, the authors  of \cite{dns} used a rather
contrived ``modular" approach, relying on multiple stages of symmetry breaking, 
generating 
thereby a perturbative
hierarchy of scales in the supersymmetry breaking sector.
In addition to being aesthetically unappealing (admittedly a rather subjective
criterion!), the MGM models also require a significant amount of fine tuning
(the most drastic example being the dimensionless 
Yukawa coupling in the ``$3-2$"
model of supersymmetry breaking, which, for calculability,
 needs to be of order $10^{-7}$). }

\item{In addition to aesthetic and fine tuning problems, the models also face the 
issue that the supersymmetry breaking vacuum is (almost) always 
local (this was pointed
out in refs.~\cite{CHL}, \cite{DDR}). It is a generic
feature of models with messenger singlets that there is  a supersymmetry
preserving, deeper ground state. To illustrate this, recall that the messenger
singlet superfield, $S$,  couples in the superpotential to the messenger
quarks and leptons, $Q, \bar{Q}$ ($G_{SM}$ indices are suppressed below):
\beq
\label{singlet}
W = \lambda~ S ~Q \cdot \bar{Q} - F_S^*~ S~.
\eeq
The last term in eq.~\ref{singlet} is added to represent
 the fact that when $\lambda = 0$,
 the
messenger singlet obtains a supersymmetry breaking ($F$-type) expectation value.
It is easy to see that when the Yukawa coupling $\lambda$ of the messenger
singlet to the messenger quarks and leptons is nonvanishing, the equation of 
motion for $S$, $\partial W/\partial S = 0$, always has a supersymmetry
preserving solution with $\langle Q \bar{Q}\rangle \ne 0$ along the $D$-flat
directions of $G_{SM}$. In addition
to preserving supersymmetry, this minimum also breaks color and
electromagnetism.}

\item{Finally, in the MGM models with messenger singlets,
gauge symmetries alone  do not forbid
 direct mass terms for the messenger quark and 
lepton superfields. Typically, one would have  
to impose additional global symmetries to prevent 
explicit mass parameters
from appearing  in eq.~\ref{singlet}.}
\end{enumerate}

Our hope is that a successful model that unifies the supersymmetry breaking
and messenger sectors will be able to alleviate the  problems   listed above.
 To illustrate why we think that {\bf 1.--3.} might be avoided, let us imagine
that the trilinear coupling in (\ref{singlet}) is replaced as follows:
\beq
\label{replace}
\lambda ~S ~Q~ \bar{Q}~\rightarrow \lambda ~[S]_{A B} ~Q^A \cdot \bar{Q}^B~.
\eeq
In writing eq.~\ref{replace}, we have assumed that the messenger quarks and
leptons now transform also in some representation of the supersymmetry
breaking gauge group $G_{SB}$ 
($A$ and $B$ denote the corresponding indices). 
The square
bracket around the field $S$ indicates that it need not be a fundamental field: 
it can be a product of fields (a composite) with the desired transformation
property under $G_{SB}$. We hope that the problems listed above can be alleviated,
since:
\begin{enumerate}
\item{If the right $G_{SB}$ is found, one could avoid the ``modular" structure
of the MGM models and thus gain elegance, and avoid fine tuning.}
\item{Since the messenger fields are now part of the supersymmetry breaking
dynamics,  $G_{SB}$, the supersymmetry breaking minimum may be global. Recall that
in  MGM models, the coupling $\lambda$ of the messenger singlet to the
messengers (\ref{singlet}) 
causes the appearance of $G_{SM}$-breaking minima, since the messengers
do not participate in supersymmetry breaking. In models where the 
messengers are an integral part of the supersymmetry
breaking sector, this coupling may in fact be required in order 
to achieve supersymmetry breaking---it may be needed to lift 
some runaway directions.}
\item{Finally, if the messengers are in chiral 
representations of the supersymmetry
breaking gauge group $G_{SB}$, direct mass terms for the messengers are forbidden
by gauge invariance alone and 
 there is no need to impose additional global symmetries.}
\end{enumerate}

\subsection{A new problem: loss of asymptotic freedom?}

\smallskip

The major problem that arises when one 
tries to unify the supersymmetry breaking
and messenger dynamics was noted some time
 ago \cite{dn93}: in the models 
that we wish
to construct the messenger fields carry quantum numbers under both 
$G_{SB}$ and $G_{SM}$. Thus, $G_{SB}$ is a ``flavor" symmetry of the
messenger sector, which implies the existence of multiple copies of 
messenger quark and lepton supermultiplets. 
Conversely, $G_{SM}$ is a  ``flavor"
symmetry of the supersymmetry breaking sector.
 
 Now recall that 
 the standard model gauge couplings are asymptotically free
 only if  no more than 
4 copies of messengers in the ${\bf 5 + 5^*}$ are added (for simplicity
we list the field content in terms of $SU(5)$ representations). Thus, if we insist on
preserving asymptotic freedom for the standard model, the representation of 
$G_{SB}$ under which the messenger quarks and leptons transform should have dimension 
less than or equal to 4. This in turn implies that the supersymmetry breaking
gauge group $G_{SB}$ should be rather ``small", i.e. have representations of
dimension $\le 4$. We conclude that if we are to keep asymptotic 
freedom of $G_{SM}$, 
we require a ``small" $G_{SB}$ (such that it has representations 
of dimension $\le 4$)
 with a ``large" enough {\it unbroken} flavor symmetry,
such that $G_{SM} \subset G_{flavor}$.

Given our limited experience (since Seiberg's work in 1994) 
in building theories that
dynamically break supersymmetry, constructing  a supersymmetry breaking sector
that has the properties listed above 
appears hard. Typically, achieving large flavor symmetries is difficult, 
without having
to add explicit mass terms. Moreover, verifying that the flavor symmetry
(part of  which is identified with $G_{SM}$) is unbroken by the supersymmetry
breaking dynamics is  presently only possible in models where the 
infrared dynamics is  calculable. We note that 
we have constrained our
considerations to the case when the supersymmetry breaking gauge group is
a simple group or, at most, $G_{SB} = G_1 \times G_2$, since the dynamics of 
 models with more than two groups in the product
is usually much harder to analyze. However, as 
A. Nelson has informed us, it may be possible to construct a satisfactory
model without Landau poles of $G_{SM}$
 if $G_{SB}$ is a product 
 of three ``small" gauge groups, 
 e.g. $G_{SB} = SU(2) \times SU(3) \times SU(4)$.

An interesting possibility, which would allow us to dispose of the requirement
of asymptotic freedom,
 is that the Landau poles of the standard model gauge couplings may 
only signal the breakdown of our description at higher energies, rather than a 
fundamental inconsistency of the theory. 
As was realized
by Seiberg \cite{seiberg}, an infrared free theory can be just 
the effective low-energy weakly coupled
description  of an ultraviolet free theory, which becomes 
strongly coupled in the infrared. The ``canonical" 
example is the $SU(N_c)$ supersymmetric QCD with
$N_c + 2 \le N_f \le 3 N_c/2$. This theory is free in the ultraviolet and 
 strongly coupled in the infrared. It has a weakly coupled description
 at low energies, given by  an infrared free, ``magnetic"
$SU(N_f - N_c)$ gauge theory---this infrared free theory
 would be the analog of the 
 standard model with a large number of 
extra vectorlike messengers.
If such a scenario is realized, the true description of the theory at short
distances could be quite different from the familiar
 $SU(3)\times SU(2)\times U(1)$. However,
at this stage of our understanding it appears hard to build explicit
 models that 
realize this
possibility, because  
Seiberg's duality is not well understood at nonzero momentum and in the presence
of supersymmetry breaking, and because there are no ``compelling" 
(i.e. simple, weakly coupled) 
duals of the standard model gauge group with extra vectorlike matter.

Another possibility, which is realized in the models discussed in Section~2., 
is that the scales of the Landau poles of the standard model gauge couplings
  are pushed up, beyond the GUT, string, or
Planck scale. One way to realize this possibility is to construct $G_{SB}$ models
with two scales, e.g. $M$ and $\Lambda$. 
In the models we will construct,  $M$ will be a fundamental scale,
which suppresses some nonrenormalizable operators, while
 $\Lambda$ is a dynamically
generated scale. The ratio $\Lambda/M$ is a small parameter, 
which will generate the 
desired hierarchy of scales in the supersymmetry breaking sector. 
Although in these models $G_{SB}$ can have a rather large rank,
causing  multiple copies of messengers to appear, the dynamics may be such---due
to the presence of the small parameter $\Lambda/M$---that
only a small number ($\le 4$, in terms of ${\bf 5 + 5^*}$ of 
$SU(5)$) of these messengers is light, while the others are heavy, of order the 
GUT scale (or heavier). Thus the appearance of the Landau poles of the standard
model gauge couplings is ``postponed" until scales where a description in
terms of a more fundamental
(string) theory becomes valid.

\subsection{What is needed to build models with dynamical messengers?}

\smallskip

From the discussion in the previous sections, 
it is clear that a model with dynamical 
messenger fields has to incorporate the following ingredients.
 One needs, first of all, a supersymmetry-breaking theory, which
has a large enough 
nonanomalous 
flavor symmetry (such that $G_{SM}$ can be embedded in it). In addition, the 
flavor group should be unbroken by the supersymmetry-breaking dynamics. Finally, in
order for the Landau poles to be pushed up, we require that the dynamics is such
that only a few (or none---see refs.~\cite{hitoshi}, \cite{cern}) 
light messenger fields appear in the spectrum.

\section{Existence proof: models based on ${\bf SU(N)\times SU(N-2)}$.}

In this section we will present the ``existence proof" for such models.
The model we construct is
based on the $SU(N)\times SU(N-2)$ theories of supersymmetry
breaking \cite{pst2}. As we discuss later,
 the models are not phenomenologically acceptable as purely 
 gauge-mediated models. However, 
they present the first example of models with dynamical messenger fields and
by studying their dynamics  in some detail, we 
will learn some general properties, which are expected to  be
shared by more realistic models.

\subsection{The tools: the ${\bf SU(N)\times SU(M)}$ theories.}

\smallskip

As we discussed in Section~1.3, in order
to construct models with dynamical messenger sectors, one
requires a supersymmetry-breaking theory  that has a large
enough anomaly-free flavor symmetry (so that the standard model
gauge group can be embedded in it). In addition, one would
like to be able to establish that the
flavor group is unbroken in the vacuum. At present, verifying
this in the presence of supersymmetry breaking is only possible
if the supersymmetry breaking dynamics is calculable, at least
in the infrared.

We have studied in some detail the construction of
 supersymmetry-breaking theories based on supersymmetric 
 QCD with gauged flavor symmetries. Following this ``Rube Goldberg"
 (see A. Nelson's contribution) approach  
 we arrive at a specific set of theories based on product 
 gauge groups. The advantage of this approach is that 
 models based on gauging
 flavor symmetries are relatively easy to analyze (since each
 of the gauge groups is supersymmetric QCD, the nonperturbative
 dynamics of which is well understood), yet the
 interplay of the gauge dynamics of the two groups gives rise
 to many new interesting nonperturbative 
 phenomena, specific for the product-group case
 \cite{pst1}.

By gauging the chiral flavor symmetries of supersymmetric 
QCD (and completing the 
representation to obtain the minimal anomaly-free content) we arrive
at the following class of gauge theories: the gauge group is 
 $SU(N) \times SU(M)$ with
a matter content consisting of a single field,
$Q$, that transforms as (\Yfund , \Yfund ) under the
gauge group, $M$ fields, $\bar{L}$, transforming as
$(\overline{\Yfund}, {\bf 1})$, and $N$ fields,  $\bar{R}$,
that  transform as  $({\bf 1}, \overline{\Yfund})$.  We note that
these theories  are  chiral---no mass terms can be added for any of the matter
fields. The detailed dynamics of these models was analyzed in refs.~\cite{pst1},
\cite{pst2}. Here we will only list the features relevant for model building.

We begin with the case $N = M$. The global vacuum 
in this  case does not break supersymmetry---the theories in fact
exhibit confinement without chiral symmetry breaking \cite{pst2}.
However, it is possible to show that there exist local minima which
break supersymmetry and preserve a large enough subgroup of the 
flavor symmetries \cite{cern}.
These local vacua  (with $N=5$) have been recently 
used  in the construction
of some interesting models with dynamical messengers \cite{cern}.

In the case $M=N-1$, it was shown in ref.~\cite{pst1} that with an
appropriate superpotential added, the  models break supersymmetry.
In addition, the vacuum can occur in the calculable regime \cite{shirman},
and a large enough global symmetry ($SU(N-2)$) can be left unbroken. 
However, in these models, the global symmetry is anomalous, and gauging
it requires adding ``spectator" multiplets to obtain anomaly-free representations.
The $SU(N)\times SU(N-1)$ models (with additional standard model representations) 
were used in ref.~\cite{berkeley} to construct some models of dynamical
supersymmetry breaking that unify the messenger and supersymmetry breaking sectors.
The dynamics of these models is similar to the ones discussed  below (for an
important distinction, see comment at the end of Section~2.3).

The $M=N-2$ case was considered in ref.~\cite{pst2}, \cite{pt1}. It was shown
that upon adding an appropriate superpotential the models with odd $N$
 break supersymmetry
with a stable ground state and preserve a global, anomaly-free,
symmetry $SP(N-3)$ (we use
$SP(2) = SU(2)$). These models will comprise the supersymmetry breaking
sector of the models we will construct below. The models with even $N$ 
 have global supersymmetric vacua.

For completeness, we note that at present we do not know whether 
there exist  stable supersymmetry-breaking vacua in the models
with $M<N-2$. The main difficulty in establishing this 
 lies in analyzing the 
classical flat directions. 

\subsection{The ${\bf SU(N)\times SU(N-2)}$ models: scales 
and messenger spectra.}

\smallskip

Given the results described above 
and the necessary conditions of Section~1.3, 
it is now straightforward to
build phenomenological models with dynamical messengers. Since the 
$ SU(N)\times SU(N-2)$ models
break supersymmetry and 
preserve an anomaly free global $SP(N-3)$ ($N$-odd), 
one can identify a subgroup of 
the global symmetry with the standard model gauge group. The fields in
the supersymmetry breaking sector then carry charges under both standard
model and supersymmetry-breaking gauge groups, $G_{SM}$ and
$G_{SB} = SU(N)\times SU(N-2)$. Embedding $G_{SM}$ into the $SP(N-3)$ global
symmetry requires $N \ge 11$ (while consistency with grand unification is only
possible for $N \ge 13$) \cite{pt1}. 

We will not describe in detail the dynamics of these models here. We will
only stress that the models realize the idea discussed in the previous section:
the Landau poles for the standard model gauge groups are ``postponed" until
after the GUT (or string scale). 
Two scales are essential for the dynamics
of these models. One scale,
$M$, is the scale that suppresses the nonrenormalizable 
term in the superpotential, 
which is needed to lift the flat directions. 
The second scale---the strong
coupling scale of the $SU(N)$ gauge group, $\Lambda$, $\Lambda \ll M$---is 
dynamically generated. Instead of using $M, \Lambda$ as the two scales, it 
is convenient to use $M, v$, where $v$ is the scale of some vacuum expectation
value and is a function of both $M$ and $\Lambda$ \cite{pt1}. 
We will see below that the  presence of two scales, 
$M$ and $v$, with a small ratio $v/M \ll 1$,  generates the desired hierarchy 
of mass scales.

The dynamics of the models is such that
at the scale $v$ ($v \ll M$)
the $SU(N-2)$ gauge group is totally broken. Below the scale of $SU(N-2)$ breaking,
the dynamics is governed by a pure $SU(N)$ gauge theory, with a dilaton and 
a number of moduli fields. Under the $SP(N-3)$  flavor  group, 
the light moduli fields transform as
 two fundamentals and three singlets.
At the scale 
\beq
\label{lambdalow}
\Lambda_L = v ~\left({v\over M}\right)^{N-5 \over 3}
\eeq
 the pure $SU(N)$ gauge theory confines
and the dynamics below this scale is governed by the moduli and dilaton only.
Moreover---and this is an important feature of these models pointed
out by 
Shirman \cite{shirman}---at energy scales below $\Lambda_L$, 
the dynamics of the moduli and dilaton fields
is described by a calculable nonlinear supersymmetric sigma model. 
This fact allows us
to compute the spectrum of the light excitations and confirm 
that the flavor symmetry
is unbroken in the supersymmetry-breaking ground state; for details, see
 \cite{pt1}. We note that the dynamics of this model is not {\it entirely}
 calculable---it is only the spectrum of the excitations lighter than $\Lambda_L$
 that is calculable; the spectrum
 of strongly coupled bound states of gauginos and 
 gauge bosons (of mass $\sim \Lambda_L$)
 of the pure $SU(N)$ gauge theory is beyond theoretical control. 
 In this sense the model differs from the known  
 calculable models of dynamical supersymmetry breaking,
 such as the ``3-2" model.
 
 As we already mentioned, the light spectrum only involves two fundamentals of
the $SP(N-3)$ flavor group. Thus, gauging the standard model gauge group
(say, in a way consistent with grand unification, requiring thus $N \ge13$),
the spectrum of light messengers consists of two pairs of
${\bf 5 + 5^*}$ representations of $SU(5)$. 
The masses of these light messengers are of order 
\beq
\label{m} 
m = v ~\left({v\over M}\right)^{N-5}~.
\eeq 
Finally, the supersymmetry breaking
scale is of order $F = m v$, i.e.
\beq
\label{F}
 \sqrt{F} = v ~\left({v\over M}\right)^{N-5 \over 2}.
\eeq

The masses of all   heavy messenger fields are of order $v$. 
Thus, they do not contribute to the
running of the standard model gauge couplings below that scale. The Landau poles
of the standard model are therefore ``postponed" until scales higher than $v$.
Beyond that scale, a more fundamental theory is expected to furnish the correct
description of the dynamics.

\subsection{Communication of supersymmetry breaking: 
the messenger ${\bf {\rm Str} M^2}$ and the soft parameters.}

\smallskip

The communication of supersymmetry breaking is expected to work as in the 
usual models of gauge mediation---since the messenger fields are also a part
of the supersymmetry breaking sector, they ``learn" about supersymmetry breaking
at tree level. The superpartners of the quarks, leptons, 
and standard model gauge
bosons acquire supersymmetry-breaking masses at loop level. Thus, one expects
that the order of magnitude of the soft masses is $\sim g^2 m/16 \pi^2$, as in
the MGM models. We note that there are contributions to the 
supersymmetry-breaking soft masses of the standard model 
superpartners at two scales: both
the heavy (of mass $v$) and light (of mass $m$) messengers contribute to the
soft parameters (the contributions of the heavy and light messengers are of
the same order of magnitude; for a discussion of the contribution of
the heavy messenger vector superfields see ref.~\cite{cern2}).

The spectrum of light messengers in these models, however, differs in one
crucial respect from the messenger spectrum of the MGM models. Since the 
dynamics of the supersymmetry-breaking-cum-messenger sector is governed by a 
complicated nonlinear sigma model, the masses of the messenger quarks and leptons
receive contributions from both the superpotential and K\" ahler potential. The
nontrivial K\" ahler potential contributions lead to a nonvanishing supertrace
of the light messenger mass matrix.
In fact, it is expected that in any model
 where the 
messengers participate in the breaking of supersymmetry, 
and the low energy dynamics can be described by a weakly coupled
nonlinear sigma model, ${\rm Str} ~M^2_{\rm mess} \ne 0$. This can be
seen by considering the  tree level supertrace mass squared sum rule \cite{WB} 
for a general nonlinear sigma 
model:
\beq
\label{sumrule}
{\rm Str} ~M^2   =  - 2 ~R_{i j^*} K^{i l^*} K^{m j^*} W_m W^*_{l^*} .
\eeq
Here we use the notations of ref.~\cite{WB}: $W_m$ is the 
gradient of the superpotential, $K^{i j^*}$ is the inverse K\" ahler metric, and 
$R_{i j^*}$ is the Ricci tensor of the K\" ahler manifold. In eq.~(\ref{sumrule})
the trace is taken over all states in the sigma model (including the messenger
singlets), however---and explicit examples confirm this---one does not 
expect a restriction of the supertrace on the space of states charged under 
a global symmetry to vanish for a K\" ahler manifold with nonzero curvature.

As a result of the nonvanishing supertrace, the light messengers' 
contribution to the soft scalar masses turns out to be logarithmically 
 divergent; see ref.~\cite{pt2}. The divergent piece is:
\beq
\label{softscalarL}
\delta_L m^2 \sim - {g^4  \over 256 \pi^4} ~
{\rm Str} M^2_{mess} ~{\rm Log} {\Lambda^2\over m^2} .
\eeq
Here $\Lambda$
 is the ultraviolet
cutoff and $m $ is the scale of the light messengers (\ref{m}), 
introduced earlier.
The logarithm is cut off by the contributions of the 
heavy messenger fields; see discussion in ref.~\cite{pt2}.
 
 The explicit calculation of the light messenger spectrum in \cite{pt1} shows
 that in the considered vacuum, ${\rm Str} M^2_{mess} > 0$. 
 Then eq.~\ref{softscalarL} implies that
 the scalar soft masses in this vacuum receive a large logarithmically
 enhanced negative contribution. Thus, the vacuum studied in ref.~\cite{pt1} 
 can only be phenomenologically relevant if there are additional 
 positive contributions 
 to the soft scalar masses that offset the log-enhanced negative contribution due
 to the positive supertrace. A detailed numerical analysis
shows that the positive contribution of the heavy messengers
is not sufficient to cancel the large 
negative logarithmic contribution. Thus, the models can only be 
viable if in addition to the gauge mediation effects, there
are also positive supergravity contributions to the soft scalar masses. 
In fact, upon examining the energy scales
in these models, with $M = M_{Planck} = 2 \times 10^{18}$ GeV, we 
find that for $N = 13 - 17$, the 
scale $v = 10^{16}$ GeV, while the supersymmetry breaking 
scale (\ref{F}) is $\sqrt{F} = 10^{10}$ GeV 
(the scale of the light messenger fields (\ref{m}) is of
order $m = 10^4$ GeV). Thus, supergravity 
contributions are of the same order
as the gauge-mediated contributions. Therefore we
 refer to these models as ``hybrid" models
of supersymmetry breaking. We note that the scalar soft masses in the hybrid models
receive contributions at three different
 energy scales---the (uncalculable) supergravity contribution
 at $M = M_{Planck}$ and the contributions from the heavy  and light messengers
 at the scales $v$ and $m$, respectively.  On the other hand,
  gauginos only receive contributions from gauge mediation,
   at the scales $v$ and $m$. One
expects that in the ``hybrid" models the 
renormalization group running will be 
different than that in pure  supergravity or gauge mediated models---
in particular, since only the squarks receive the negative supertrace
contribution, it might be possible to obtain soft-parameter 
spectra with squarks and sleptons lighter than the gauginos.

Finally, we would like to stress that the scalar potential of the nonlinear
sigma model that describes the infrared dynamics of
 the $SU(N)\times SU(N-2)$ models is very complicated---we only studied the 
 simplest solution---and it is not excluded that there exist vacua where the
 supertrace has a negative sign. We also note that a negative supertrace 
 can not occur in the
models  \cite{berkeley}  based on $SU(N)\times SU(N-1)$. This is 
essentially due to 
the fact that, as we mentioned in Section~2.1, the global $SU(N-2)$
symmetry is anomalous. 't Hooft
anomaly matching then implies the existence of a 
massless fermion in the supersymmetry breaking
sector, leading to a positive supertrace of the light messengers. 
The addition of extra matter to form anomaly-free representations can 
give a Dirac mass to this fermion \cite{berkeley}, which is, however, a 
small  higher-order effect and can not change the sign of the supertrace.

If models with dynamical messengers---or new vacua of the existing models---are
 found, with negative messenger supertrace, one expects to obtain a
 soft spectrum different than the one in the MGM models, with the scalar soft
 masses logarithmically enhanced with respect to the gaugino masses (we note that
 some models that have negative supertrace were constructed in refs.~\cite{lisa},
 \cite{bogdan}). The phenomenological viability of such models deserves a
 closer investigation.

\section{What have we learned?}

In the course of this investigation, we showed that models with dynamical 
messengers can be constructed. We outlined the necessary requirements to build
successful such models, and proposed a solution to the Landau pole problem,
where the Landau poles are ``postponed" to higher energy scales---beyond the 
GUT or string scales---where a more fundamental description of the dynamics
is expected to take over.

We constructed an explicit example of such models, based on the $SU(N) \times
SU(N-2)$ theories of dynamical supersymmetry breaking. We showed that the 
supersymmetry breaking dynamics leaves a large enough global symmetry
unbroken, and calculated the detailed spectrum of light  messenger fields. 

We showed that, quite generally, in models with dynamical messenger fields, 
where the low-energy dynamics is described by a weekly coupled nonlinear sigma 
model, one expects the supertrace of the light messenger fields to 
be nonvanishing. We showed 
that this supertrace leads to logarithmically enhanced contributions to the
soft scalar masses. 
Positivity of the soft scalar mass squared requires generally that
the sign of the supertrace of the light messengers is negative.

Clearly, a lot of work remains to be done to find the "Standard Model"
of gauge-mediated supersymmetry breaking (for some interesting new
developments,
see the contribution of H. Murayama to this volume and 
refs.~\cite{hitoshi}, \cite{cern}).
As we previously emphasized,
although the models constructed here are not acceptable
 as purely gauge-mediated
models, but can only work as ''hybrid" models, 
the detailed investigation of the  supersymmetry-breaking dynamics
and generation of the soft masses teaches us some general properties of
such models. The lessons learned 
are likely to be useful in building models that are more aesthetically
appealing and phenomenologically viable.

\nc{\ib}[3]{ {\em ibid. }{\bf #1} (19#2) #3}
\nc{\np}[3]{ {\em Nucl.\ Phys. }{\bf #1} (19#2) #3}
\nc{\pl}[3]{ {\em Phys.\ Lett. }{\bf #1} (19#2) #3}
\nc{\pr}[3]{ {\em Phys.\ Rev. }{\bf #1} (19#2) #3}
\nc{\prep}[3]{ {\em Phys.\ Rep. }{\bf #1} (19#2) #3}
\nc{\prl}[3]{ {\em Phys.\ Rev.\ Lett. }{\bf #1} (19#2) #3}
\nc{\ptp}[3]{ {\em Progr.\ Theor.\ Phys.}{\bf #1} (19#2) #23}

\end{document}